\documentstyle[budapest1]{article}  
\frompage{000} \topage{000}                                              
\def\rightmark{Faces of quark matter} 
\def\leftmark{J. Zim\'anyi et al.}
 
\newcommand{\insertplotg}[1]{\begin{center}\leavevmode\epsfysize=10.5cm 
\epsfbox{#1}\end{center}}

\title{Faces of quark matter} 
\authors{
{J. Zim\'anyi, P. L\'evai, T.S. Bir\'o }\\[2.812mm]
{\normalsize
Research Institute for Particle and Nuclear Physics \\
POB. 49. Budapest, 1525, Hungary \\[0.2ex] 
}}
 
\abstract{
Based on an analysis in the framework of a coalescence hadronization
model (ALCOR) we conclude that in heavy ion collisions at CERN SPS and RHIC energies
a new type of matter, the massive quark-antiquark matter is produced.
}

\keyword{quark-gluon plasma, hadronization} 
\PACS{25.75.-q}
 
\begin{document}
 
\maketitle
\setcounter{page}{1}

\section{Introduction}\label{intro}

The heavy ion community is on the search of quark-gluon plasma already for
more than 30 years. For this purpose large amount of money was spent. Thus
it is timely to ask the question: what have we delivered in exchange
for this support? Could we produce a piece of quark-gluon plasma?
 Have we determined the properties of this matter? Could we arrive to
 results which will be worthwhile to mention after a few years
has passed?

  In this lecture we try to shed some light to a few of these questions. 

\section{Time evolution of heavy ion collision}\label{qualitative}

The different reaction models offer even qualitatively very
 dissimilar pictures for the
flow chart of a heavy ion collision (see Fig.1).
Some of the models assume that the first collision of the incoming
nucleons produce directly most of the hadrons appearing in the final state. 
In contrast  other models are based on the  production of a collective 
intermediate state which expands and hadronize in a later stage only.
Some of them use even the concept of thermal equilibrized intermediate
state. The main goal of heavy ion research is to find out
whether this collective state is formed in the heavy ion collision,
and if it is formed then what are its properties.

\newpage
\begin{figure}[tp]
\insertplotg{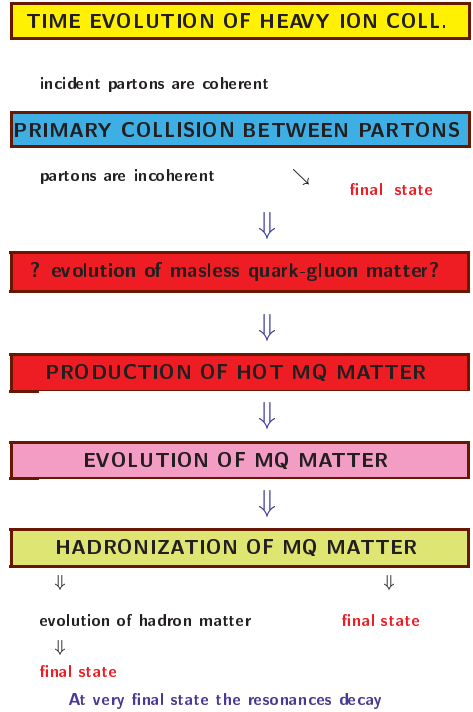}
\vspace*{-0.5cm}
\caption{The flowchart of the heavy ion collision. Notation: 
MQ = massive quark}
\label{flow}
\end{figure}

\newpage

\section{Qualitative considerations}

When we try to draw a picture of the heavy ion collisions,
it  is enlightening to recall two earlier ideas (perhaps in a 
somewhat subjective way).

V.N. Gribov pointed out that in the nucleon-nucleus reaction at the
first collision the incident nucleon looses its low momentum 
components, since their cross section are the largest (Ref.~\cite{Gribov}).
The remaining "wounded nucleon" (A. Bialas in Ref.~\cite{Bial1}) will interact
with the next nucleon in a modified way.

According to the above considerations we can distinguish the
 following type of collisions:

 a) soft collisions: the dress of both incoming nucleons is
 intact;

 b) semi-hard collisions: one of the incoming nucleons is "wounded", 
while the other incoming nucleon has an intact dressing;

 c) both incoming nucleons are  wounded.

\vspace{4mm}

One more fact: in a recent calculation it was shown that
 the transverse momentum distribution of partons produced in the
collision of bombarding and target partons has an exponential
shape \cite{Levai}. 
Fig. 2 displays the $m_t$ distributions of the produced massive quarks 
obtained in a perturbative QCD calculation,
The spectral shape, which is exponential in a wide $ m_t $ range
is a result of the structure of the
reaction matrix element. Thus the observation of an exponential
spectrum is by no means an indication of any thermalization.

\vspace{4mm}

With these considerations in mind we may phrase the following 
problems:

\noindent
$\bullet$ Is it true, that the strange quark-antiquark pairs ($s{\overline s}$
are produced thermally in the fireball?

\noindent
$\bullet$ Or are they produced in the primary collisions?

\noindent
$\bullet$ In the latter  case the $ s \overline s $ pairs
 are not a signature for 
quark-antiquark plasma.

\vspace{4mm}

Nevertheless the free recombination of massive quarks and antiquarks 
is a signature for
the formation of deconfined quark-antiquark matter.

\begin{figure}[htbp]
\setlength{\epsfxsize=0.55\textwidth}
\setlength{\epsfysize=0.45\textheight}
\centerline{\epsffile{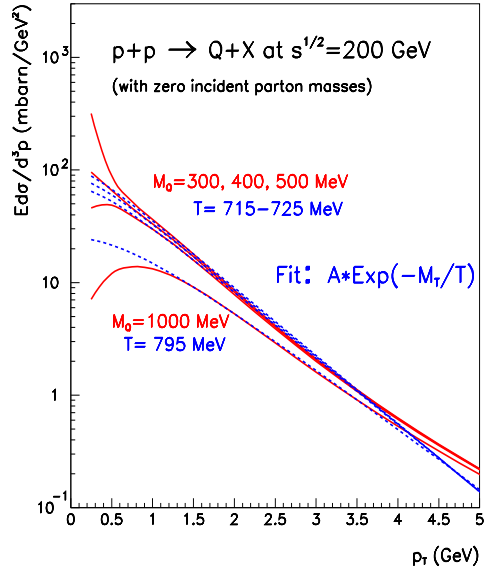}}
\caption[]{
Invariant transverse momentum spectra of
massive quarks  produced in $p+p$ reaction at
$\sqrt{s}=200$ GeV. The incident parton masses are zero, the
produced quarks are heavy with effective masses  $M_Q = 300,400,500,1000$ MeV.
Figure is redrawn from Ref.~\cite{Levai}.
}
\label{figure2}
\end{figure}

\newpage

\section{{The early concept of the quark matter }}

At the beginning for the description of hadron matter --- quark-gluon plasma
phase transition we used the following ideas:

\noindent
$\bullet$ at the beginning of heavy ion collision a quark-gluon 
plasma with massless quarks, antiquark and gluons is produced; 

\noindent
$\bullet$ a first order phase transition occurs, quasistatically 
satisfying the Gibbs criteria; 

\noindent
$\bullet$ homogeneous distributions and a few averaged
particle properties (e.g. $ \mu_q $, $\mu_s$, T) describe everything

\noindent
 ALL THESE IDEAS ARE KILLED BY THE NEW OBSERVATIONS 
obtained at CERN SPS for the rapidity dependence of the
multiplicity ratios (see Fig. 3). 

\vspace*{-2truecm}
\begin{figure}[hp]
\setlength{\epsfxsize=0.45\textwidth}
\setlength{\epsfysize=0.45\textheight}
\centerline{\epsffile{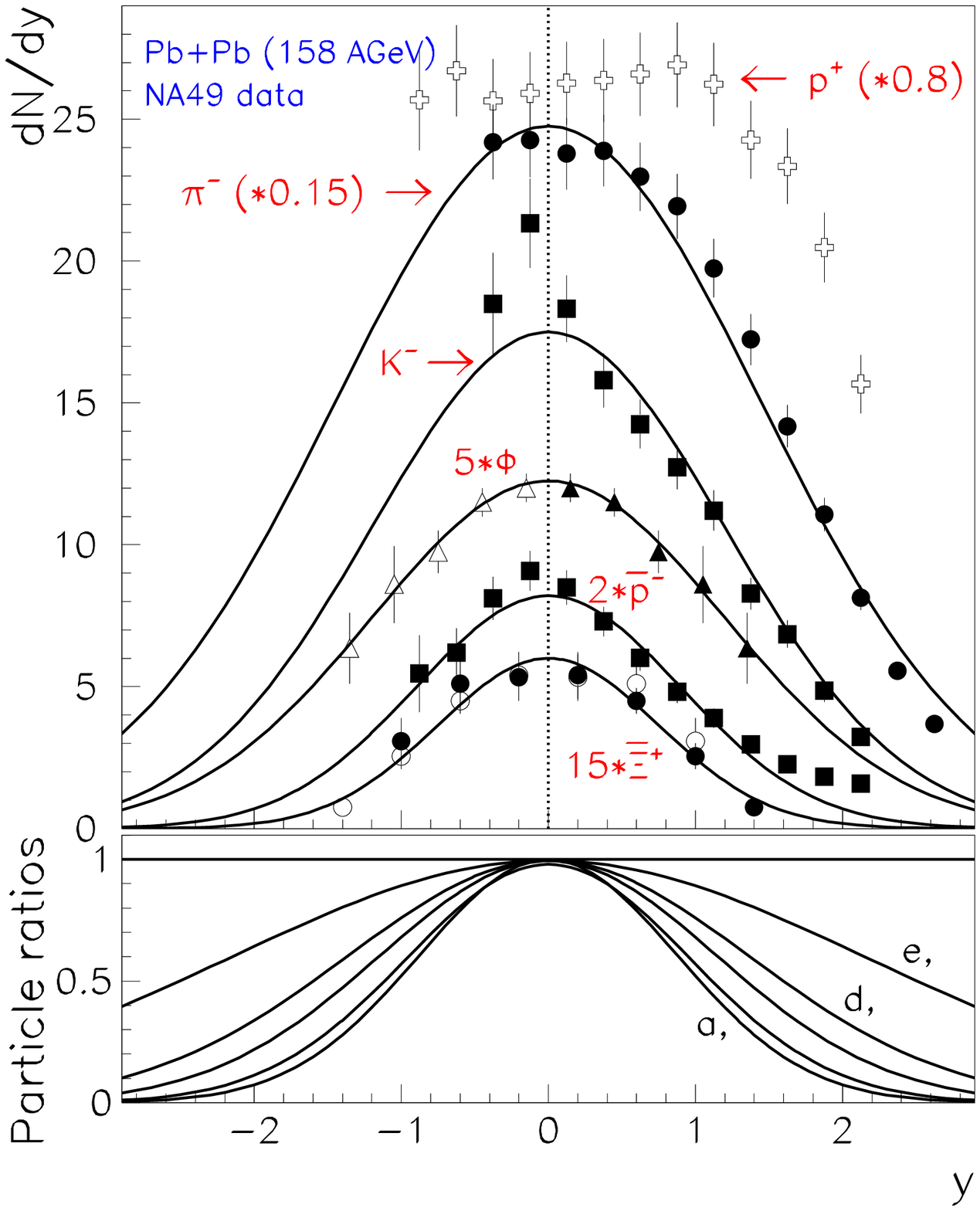}}
\vspace*{-0.4truecm}
\caption[]{
Top panel: Normalized rapidity spectra of hadrons produced in
Pb+Pb collision at $E_{LAB}=158$ AGeV.
Data are from Refs.\cite{NA49,NA49SQM99}.
Bottom panel:
particle ratios normalized to unity at the mid-rapidity.
The lines indicate: a, ${\overline \Xi}^+/K^-$; b, ${\overline \Xi}^+/\Phi$;
c, ${\overline p}^-/\pi^-$; d, ${\overline p}^-/K^-$;
e, $\Phi/\pi^-$.}
\label{figure3}
\end{figure}
\noindent 
The BRAHMS experiment at RHIC confirms the strong rapidity dependence,  
the proton to antiproton ratio was  measured at different rapidities as:
 $ {\overline p}/p = 0.64 \pm 0.03$ at $y=0$ and 
 $ {\overline p}/p = 0.41 \pm 0.03$ at $y=2$ \cite{BRAHMS}.

It was observed also, that the expansion of the hot region
is very fast, excluding
the possibility of a slow quasistatic phase transition.
 
What is the way out from this contradictory situation?

{\bf One has to introduce a new concept instead of assuming an ideal 
massless quark-gluon plasma.
 Such an attempt is  done in the ALCOR model introducing the
massive quark-antiquark matter.}

\newpage

\section{The   ALCOR (ALgebraic COalescence Rehadronization)  \\ model }

In the ALCOR model \cite{ALC95,ALC99}
we assume, that just before the hadronization the dense matter
can be described as a mixture of dressed up, massive quarks and
antiquarks.  The effective mass of the gluons at this point is
much larger than that of the quarks \cite{LH}, consequently the
 gluon fission into
quark-antiquark pairs is enhanced and massive gluons 
disappear from the mixture. During the hadronization
the quark numbers, as well as the antiquark numbers, remain constant.

At this point it is proper to insert a short explanation. 
At very high temperature, say at $ T > 3 * T_c $ we may assume
a non interacting quark-antiquark-gluon plasma with zero
mass (current mass) constituents. However, as we approach the 
critical temperature, the interaction between the constituents
 will become stronger and stronger. This intensive interaction
is expressed by the large effective mass of the constituents, which
 automatically reduce the pressure, faster then the energy density.

\vspace{2mm} 
\subsection{ Basic equations of the ALCOR hadronization model}
\vspace{2mm}

According to the coalescence assumption the hadron
 numbers are proportional to the number of quarks from which 
they consist. For baryons and antibaryons we assume
\begin{eqnarray}
    {p}  &=& C_{\rm p} \,  b_{\rm q} \,  b_{\rm q}
    \,  b_{\rm q} \,  q \,  q \,  q
\hspace*{1.2 truecm}
 {\overline p}  = C_{\overline p} \,
 b_{\overline {\rm q}} \,  b_{\overline {\rm q}} \,
 b_{\overline {\rm q}} \,
 {\overline q} \,  {\overline q} \,  {\overline q}
\hspace*{1.2 truecm}
 \pi  = C_\pi \,
 b_{\rm q} \,  b_{\overline {\rm q}} \,  q \,  {\overline q}
\nonumber \\
    {\Lambda} &=& C_{\Lambda} \,
    b_{\rm q} \,  b_{\rm q} \,  b_{\rm s} \,  q \,  q \,  s
\hspace*{1.2 truecm}
{\overline \Lambda}  = C_{\overline \Lambda} \,
b_{\overline {\rm q}} \,  b_{\overline {\rm q}} \,
b_{\overline {\rm s}} \,
{\overline q} \,  {\overline q} \,  {\overline s}
\hspace*{1.2 truecm}
    K  = C_K \,
b_{\rm q} \,  b_{\overline {\rm s}} \,  q \,  {\overline s}
\nonumber \\
    {\Xi}  &=& C_{\Xi} \,
    b_{\rm q} \,  b_{\rm s} \,  b_{\rm s} \,  q \,  s \,  s
\hspace*{1.2 truecm}
{\overline \Xi}  = C_{\overline \Xi} \,
b_{\overline {\rm q}} \,  b_{\overline {\rm s}} \,
b_{\overline {\rm s}} \,
{\overline q} \,  {\overline s} \,  {\overline s}
\hspace*{1.2 truecm}
    {\overline K}  = C_{\overline K} \,
b_{\overline {\rm q}} \,  b_{\rm s} \,  {\overline q} \,  s
\nonumber \\
    {\Omega}  &=& C_{\Omega} \,
    b_{\rm s} \,  b_{\rm s} \,  b_{\rm s} \,  s \,  s \,  s
\hspace*{1.2 truecm}
 {\overline \Omega}  = C_{\overline \Omega} \,
 b_{\overline {\rm s}} \,  b_{\overline {\rm s}} \,
 b_{\overline {\rm s}} \,
 {\overline s} \,  {\overline s} \,  {\overline s}
\hspace*{1.2 truecm}
    {\eta} = C_{\eta} \,
  b_{\overline {\rm s}} \,  b_{\rm s} \,  {\overline s} \,  s
\end{eqnarray}

Here the normalization coefficients, $b_{\rm i}$,  are determined uniquely
by the requirement, that
  the number of the massive quarks do not change during the hadronization.
Due to this principle
 we have {4} ( {6} , if u and d are treated separately) conserved
 quantities  in contrast to the 2 parameters {$ \mu_{baryon},
 \mu_{strange} $} of the thermal models

This is the basic assumption for all quark counting methods
The equations showing that the number of a given type
of quarks in the final hadron population must be equal to the
number of this type of quarks in the hadronizing quark matter
are as follows:   
\begin{eqnarray}
 s &=& 3 \,  \Omega + 2 \,  \Xi +  \Lambda +
       {\overline K} +  {\eta}  \nonumber \\
 {\overline s}  &=& 3 \,  {\overline \Omega} +
2 \,  {\overline \Xi} +   {\overline \Lambda} +
        {K} +  {\eta} \nonumber \\
 q &=& 3 \,  p +  \Xi + 2 \,  \Lambda +  {K}+
       \pi  \nonumber \\
 {\overline q}  &=& 3 \,  {\overline p} +
  {\overline \Xi} + 2 \,  {\overline \Lambda} +
        {\overline K} +  \pi  \ . \label{cons}
\end{eqnarray}

\subsection{ The consequences of the simplified equations}

 Applying the trivial assumption that the coalescence 
coefficient for a given type of particle and antiparticle is the
 same : $ C_x = C_{\overline x} $ , we obtain the following antibaryon 
to baryon ratios: 

\begin{eqnarray}
   \frac{\overline p}{p}  &=& {
{ b_{\overline {\rm q}} \,  b_{\overline {\rm q}} \,
b_{\overline {\rm q}} \,
{\overline q} \,  {\overline q} \,  {\overline q} }
\over
{  b_{\rm q} \,  b_{\rm q}
    \,  b_{\rm q} \,  q \,  q \,  q }
 }
\nonumber \\
\frac{\overline \Lambda}{\Lambda} &=& { {
 {b_{\overline {\rm q}} \,  b_{\overline {\rm q}} \,
b_{\overline {\rm s}} } \,
{\overline q} \,  {\overline q} \,  {\overline s}  }
\over
 {b_{\rm q} \,  b_{\rm q} \,  b_{\rm s} \,  q \,  q \,  s } } =
{{  b_{\rm q} \,  b_{\overline {\rm s}} \, q \, {\overline s} }
\over
 { b_{\overline {\rm q}} \, b_{{\rm s}} \,{\overline q} \,
 s \,}} \cdot  \frac{\overline p}{p}
\nonumber \\
\frac{\overline \Xi}{\Xi} &=& { {
 {b_{\overline {\rm q}} \,  b_{\overline {\rm s}} \,
b_{\overline {\rm s}} } \,
{\overline q} \,  {\overline s} \,  {\overline s}  }
\over
 {b_{\rm q} \,  b_{\rm s} \,  b_{\rm s} \,  q \,  s \,  s } }
= {{  b_{\rm q} \,  b_{\overline {\rm s}} \, q \, {\overline s} }
\over
 { b_{\overline {\rm q}} \, b_{{\rm s}} \,{\overline q} \,
 s \,}} \cdot  \frac{\overline \Lambda}{\Lambda}
\nonumber \\
\frac{\overline \Omega}{\Omega} &=& { {
 {b_{\overline {\rm s}} \,  b_{\overline {\rm s}} \,
b_{\overline {\rm s}} } \,
{\overline s} \,  {\overline s} \,  {\overline s}  }
\over
 {b_{\rm s} \,  b_{\rm s} \,  b_{\rm s} \,  s \,  s \,  s } }
= {{  b_{\rm q} \,  b_{\overline {\rm s}} \, q \, {\overline s} }
\over
 { b_{\overline {\rm q}} \, b_{ {\rm s}} \,{\overline q} \,
 s \,}} \cdot  \frac{\overline \Xi}{\Xi}
\label{simp1}
\end{eqnarray}
 
Here we introduce the notation
\begin{eqnarray}
  D = {{  b_{\rm q} \,  b_{\overline {\rm s}} \, q \, {\overline s} }
\over
 { b_{\overline {\rm q}} \, b_{{\rm s}} \,{\overline q} \,
 s \,}} = \frac{K}{\overline K}
\end{eqnarray}

and thus with this we can 
 can write eq.~(\ref{simp1}) in the simple form

\begin{equation}
\frac{\overline \Lambda}{\Lambda} = D \cdot  \frac{\overline p}{p}
\hspace*{1truecm}
\frac{\overline \Xi}{\Xi} = D \cdot  \frac{\overline \Lambda}{\Lambda}
\hspace*{1truecm}
\frac{\overline \Omega}{\Omega} = D \cdot  \frac{\overline \Xi}{\Xi}
\label{simp2}
\end{equation}
        
Such simple relations as shown in eqs.~(\ref{simp1})-(\ref{simp2}) were
 first obtained for the
linear coalescence model ( $b_i \equiv 1 $ for all $b_i$ )  in
 Ref.~\cite{Bial98}. 

\subsection{  Coalescence equations without isospin symmetry}

The effect of isospin asymmetry also can be taken into account in ALCOR
by the proper quark combinatorics in the rehadronization. After
  relaxing the isospin symmetry assumption we  treat 
the $ u $ and $ d $ quarks separately.

 The coalescence equations for baryonic and mesonic flavor clusters
now read: 
\begin{eqnarray}
    {p}^+  &=& C_{\rm p} \,  b_{\rm u} \,  b_{\rm u}
    \,  b_{\rm d} \,  u \,  u \,  d
\hspace*{1.5 truecm}
    {\pi}^+  = C_{\pi^+} \,
b_{\rm u} \,  b_{\overline {\rm d}} \,  u \,  {\overline d}
\nonumber \\
    {n}^0  &=& C_{\rm n} \,  b_{\rm u} \,  b_{\rm d}
    \,  b_{\rm d} \,  u \,  d \,  d
\hspace*{1.6 truecm}
    \pi^0  = C_{\pi^0} \,
( b_{\rm u} \,  b_{\overline {\rm u}} \,  u \,  {\overline u}
+ b_{\rm d} \,  b_{\overline {\rm d}} \,  d \,  {\overline d} )/2
\nonumber \\
    {\Lambda}^0 + {\Sigma^0} &=& C_{\Lambda} \,
    b_{\rm u} \,  b_{\rm d} \,  b_{\rm s} \,  u \,  d \,  s
\hspace*{1.53 truecm}
    \pi^-  = C_{\pi^-} \,
b_{\rm d} \,  b_{\overline {\rm u}} \,  d \,  {\overline u}
\nonumber \\
    {\Sigma}^+ &=& C_{\Sigma^+} \,
    b_{\rm u} \,  b_{\rm u} \,  b_{\rm s} \,  u \,  u \,  s
\hspace*{1.25 truecm}
    K^+  = C_{K^+} \,
b_{\rm u} \,  b_{\overline {\rm s}} \,  u \,  {\overline s}
\nonumber \\
    {\Sigma}^- &=& C_{\Sigma^-} \,
    b_{\rm d} \,  b_{\rm d} \,  b_{\rm s} \,  d \,  d \,  s
\hspace*{1.25 truecm}
    K^{-}  = C_{ K^{-}} \,
b_{\overline {\rm u}} \,  b_{\rm s} \,  {\overline u} \,  s
\nonumber \\
    {\Xi}^0  &=& C_{\Xi^0} \,
    b_{\rm u} \,  b_{\rm s} \,  b_{\rm s} \,  u \,  s \,  s
\hspace*{1.38 truecm}
    K^0  = C_{K^0} \,
b_{\rm d} \,  b_{\overline {\rm s}} \,  d \,  {\overline s}
\nonumber \\
    {\Xi^{-}} &=&  C_{\Xi^{-}} \,
    b_{\rm d} \,  b_{\rm s} \,  b_{\rm s} \,  d \,  s \,  s
\hspace*{1.35 truecm}
    {\overline K^0}  = C_{\overline K^0} \,
b_{\overline {\rm d}} \,  b_{{\rm s}} \, {\overline {\rm d}} \,  \,  s
\nonumber \\
    {\Omega}^-  &=& C_{\Omega} \,
    b_{\rm s} \,  b_{\rm s} \,  b_{\rm s} \,  s \,  s \,  s
\hspace*{1.85 truecm}
    {\eta} = C_{\eta} \,
b_{\overline {\rm s}} \,  b_{\rm s} \,  {\overline s} \,  s
\nonumber \\
&& 
\hspace*{4truecm}
    {\eta}'  = C_{{\eta}'} \,
( b_{\rm u} \,  b_{\overline {\rm u}} \,  u \,  {\overline u}
+ b_{\rm d} \,  b_{\overline {\rm d}} \,  d \,  {\overline d} )/2
\label{coal2a}
\end{eqnarray}

Here again the normalization coefficients, $b_{\rm i}$,  are determined uniquely
by the quark number conservation demand:
\begin{eqnarray}
 u &=& 2~p^+ + n^0 +  \Lambda^0 + \Sigma^0 + 2~\Sigma^+ + \Xi^0  +  
       \pi^+ + (\pi^0 + {\eta}' )/2  +  {K}^+  \nonumber \\
 {\overline u}  &=& 2~{\overline p}^- + {\overline n}^0 +
   {\overline \Lambda}^0 + {\overline \Sigma}^0 +
 2~{\overline \Sigma}^- + {\overline \Xi}^0 +
     \pi^- + (\pi^0 + {\eta}')/2 +   K^-  \nonumber  \\
 d &=& 2~n^0 + p^+ +  \Lambda^0 +  \Sigma^0 + 2~\Sigma^- +   \Xi^- + 
     \pi^- + (\pi^0 + {\eta}')/2 +  {K}^0  
         \nonumber \\
 {\overline d}  &=& 2~{\overline n}^0 + {\overline p}^- +
    {\overline \Lambda}^0 +  {\overline \Sigma}^0 +
 2~{\overline \Sigma}^+ +  {\overline \Xi}^+
    +  \pi^+ + (\pi^0 + {\eta}')/2 +     {\overline K}^0
    \nonumber \\ 
 s &=& 3~\Omega^- + 2~\Xi^- + 2~\Xi^0 +  \Lambda^0 + 
       \Sigma^0 + \Sigma^+ + \Sigma^- +
       {\overline K}^0 + K^- +  {\eta}  \nonumber \\
 {\overline s}  &=& 3~{\overline \Omega}^+ +
 2~{\overline \Xi}^+ + 2~{\overline \Xi}^0 +   
   {\overline \Lambda}^0 + {\overline \Sigma}^0 +
 {\overline \Sigma}^- + {\overline \Sigma}^+ + {K}^0 +K^+ + {\eta} 
\end{eqnarray}
  \label{cons1}
In eq.~(10) $\pi $ is the number of directly produced pions.
(Note that most of the observed pions are created in the decay
of resonances.)

\subsection{ Consequences of the isospin asymmetric equations }

The isospin asymmetric ALCOR gives the following baryon ratios:

\begin{eqnarray}
   \frac{\overline p}{p}  &=& {
{ b_{\overline {\rm u}} \,  b_{\overline {\rm u}} \,
b_{\overline {\rm d}} \,
{\overline u} \,  {\overline u} \,  {\overline d} }
\over
{  b_{\rm u} \,  b_{\rm u}
    \,  b_{\rm d} \,  u \,  u \,  d }
 }
\nonumber \\
\frac{\overline \Lambda}{\Lambda} &=& { {
 {b_{\overline {\rm u}} \,  b_{\overline {\rm d}} \,
b_{\overline {\rm s}} } \,
{\overline u} \,  {\overline d} \,  {\overline s}  }
\over
 {b_{\rm u} \,  b_{\rm d} \,  b_{\rm s} \,  u \,  d \,  s } } =
{{  b_{\rm u} \,  b_{\overline {\rm s}} \, u \, {\overline s} }
\over
 { b_{\overline {\rm u}} \, b_{{\rm s}} \,{\overline u} \,
 s \,}} \cdot  \frac{\overline p}{p}
\nonumber \\
\frac{\overline \Xi^0}{\Xi^0} &=& { {
 {b_{\overline {\rm s}} \,  b_{\overline {\rm s}} \,
b_{\overline {\rm s}} } \,
{\overline u} \,  {\overline s} \,  {\overline s}  }
\over
 {b_{\rm u} \,  b_{\rm s} \,  b_{\rm s} \,  u \,  s \,  s } }
= {{  b_{\rm d} \,  b_{\overline {\rm s}} \, d \, {\overline s} }
\over
 { b_{\overline {\rm d}} \, b_{{\rm s}} \,{\overline d} \,
 s \,}} \cdot  \frac{\overline \Lambda}{\Lambda}
\nonumber \\
\frac{\overline \Xi^+}{\Xi^-} &=& { {
 {b_{\overline {\rm d}} \,  b_{\overline {\rm s}} \,
b_{\overline {\rm s}} } \,
{\overline d} \,  {\overline s} \,  {\overline s}  }
\over
 {b_{\rm d} \,  b_{\rm s} \,  b_{\rm s} \,  d \,  s \,  s } }
= {{  b_{\rm u} \,  b_{\overline {\rm s}} \, u \, {\overline s} }
\over
 { b_{\overline {\rm u}} \, b_{{\rm s}} \,{\overline u} \,
 s \,}} \cdot  \frac{\overline \Lambda}{\Lambda}
\nonumber \\                                                           
\frac{\overline \Omega}{\Omega} &=& { {
 {b_{\overline {\rm s}} \,  b_{\overline {\rm s}} \,
b_{\overline {\rm s}} } \,
{\overline s} \,  {\overline s} \,  {\overline s}  }
\over
 {b_{\rm s} \,  b_{\rm s} \,  b_{\rm s} \,  s \,  s \,  s } }
= {{  b_{\rm d} \,  b_{\overline {\rm s}} \, d \, {\overline s} }
\over
 { b_{\overline {\rm d}} \, b_{{\rm s}} \,{\overline d} \,
 s \,}} \cdot  \frac{\overline \Xi^+}{\Xi^-}
\label{simp2a}
\end{eqnarray}
  
Let us introduce the notations:
\begin{equation}
  B = {{  b_{\rm d} \,  b_{\overline {\rm s}} \, d \, {\overline s} }
\over
 { b_{\overline {\rm d}} \, b_{{\rm s}} \,{\overline d} \,
 s \,}} = \frac{K^0}{\overline K^0} \ ,
\hspace*{0.6truecm}
  D = {{  b_{\rm u} \,  b_{\overline {\rm s}} \, u \, {\overline s} }
\over
 { b_{\overline {\rm u}} \, b_{{\rm s}} \,{\overline u} \,
 s \,}} = \frac{K^+}{ K^-} \ ,
\hspace*{0.6truecm}
  C = { B \over D } = { {K^0 / {\overline K^0}} \over { K^+ / K^-} }
\end{equation}       

Using these factors B, C and D, antibaryon to baryon ratios can be written as
\begin{eqnarray}
\frac{{\overline \Lambda}^0}{\Lambda^0} &=& D \cdot \frac{{\overline p}^-}{p^+}
\hspace*{1 truecm}
\frac{\overline \Xi^0}{\Xi^0} =
C \cdot D \cdot \frac{{\overline \Lambda}^0}{\Lambda^0}
\nonumber \\
\frac{\overline \Xi^+}{\Xi^-} &=& D \cdot \frac{\overline \Lambda^0}{\Lambda^0}
\hspace*{1truecm}  
\frac{\overline \Omega^+}{\Omega^-} =
    C \cdot D \cdot \frac{\overline \Xi^+}{\Xi^-}
\label{simp3}
\end{eqnarray}

The correction factor $ C $ shows the effect of isospin.
For a system with equal number of neutrons and protons $ C=1$.
For other cases it can be obtained by inserting the experimental
values into the defining equation, eq.~(\ref{simp3}), or calculating it
from a model.
In  the {ALCOR} 
 {$ C = 1.0 $} due to isospin  symmetry.

Addendum: in the complete ALCOR calculation the resonances 
belonging to the two lowest baryon multiplets and to the 
lowest two meson multiplets are also included. 
 
\section{ Hadron ratios at midrapidity at RHIC }
\vspace{4mm}
If we are interested of particle multiplicities in a 
restricted rapidity interval, we have to use in the equations
described above the quark numbers belonging to this interval
and then the hadron numbers obtained from the coalescence
equations are also those belonging to this interval.

\vspace{4mm}

\section{ Comparison with  experimental data }

\subsection{ Parameter free predictions of ALCOR}

The set of relations eq.~(\ref{simp1}) and eq.~(\ref{simp3}) are parameter free 
predictions. They agree surprisingly well with the 
experimental results at SPS and RHIC energies 
on the ratios \  ${\overline p}/p$, \ ${\overline \Lambda}/\Lambda$,
\ ${\overline \Xi}/\Xi$, \ ${\overline \Omega}/\Omega$ 
and $K^-/K^+$, as it was shown in Ref.~\cite{Zim99}.

\vspace{4mm}

\subsection{  The full ALCOR calculation }
\vspace{4mm}

For the full calculation we have to define the three input parameters
of ALCOR~\cite{ALC95,ALC99}:

\noindent
$\bullet$ number of new $ u {\overline u} $ and $d {\overline d}$
 pairs is obtained from the   number of $ \pi^-$ (or $h^-$);

\noindent
$\bullet$ number of new strange $s {\overline s}$ pairs  
is obtained from the  number of $ K^-$ (or $K^-/\pi^-$);  

\noindent
$\bullet$  the  stopping factor (i.e. the percentage of incident quarks
 stopped  into the investigated rapidity interval)
is obtained  from  the   ratio $ K^+/K^-$.

Now one can use the ALCOR model for Au+Au collision at
$\sqrt{s}=130$ AGeV.
We choose 211 newly produced $u{\overline u}$ and $d{\overline d}$
pairs, and 105.5 $s{\overline s}$ pairs
(which means $f_s = 
2 N_{s{\overline s}}/(N_{u{\overline u}}+N_{d{\overline d}}) =
0.5$)
 and assume  that 5.8 \%
of the total nucleon numbers stopped into the central unit of rapidity
(which means 9 protons and
14 neutrons) in the Au+Au collision. We can calculate the  rapidity densities and
hadronic ratios in the central rapidity region, where 
experimental data exist. Table 1 displays 
published experimental data and the appropriate calculated ALCOR results.
The ALCOR parameters were fixed from the first 3 measured data.

\begin{table}
\begin{center}
\begin{tabular}{llll}
\hline
&ALCOR  & Experimental  & STAR  \\
&model  & data          & reference \\ \hline\hline
 $h^- $ 
& 280        & $280 \pm 20$           & [12]   \\  
 ${K^-}/{\pi^-} $ 
& 0.14       & $0.15 \pm 0.01$        & [13]   \\
 ${K^+}/{K^-} $ 
& 1.14       & $1.14 \pm 0.06$        & [13]   \\ \hline
 ${\overline p}^-/{p}^+$
& 0.63       & $0.60 \pm 0.03$        & [13]   \\
 ${\overline \Lambda}/{\Lambda}$
& 0.73       & $0.73 \pm 0.03$        & [13]   \\
 ${\overline \Xi}^+/{\Xi}^-$  
& 0.83       & $0.82  \pm 0.08$       & [13]   \\
 ${\Xi^-}/{\pi^-} $
& 0.015      & $0.013 \pm 0.01$       & [14]   \\ 
 $\Phi/K^{*0}$
& 0.38       & $0.49 \pm 0.13$        & [15]   \\
 ${\overline p}^{-}/h-$
& 0.087     & $0.070 \pm 0.002$       & [16]   \\ \hline
\end{tabular}
\end{center}
\caption{
Hadron production in Au+Au collision at $\sqrt{s}=130$ AGeV
from the ALCOR model and the experimental data from STAR
\cite{STARh,STARc,STARv,STARk,STARap}. 
The first 3 rows are the inputs for fit the 3 necessary ALCOR parameters.}
\end{table}
 
\begin{center}
\vspace*{8.1cm}
\includegraphics{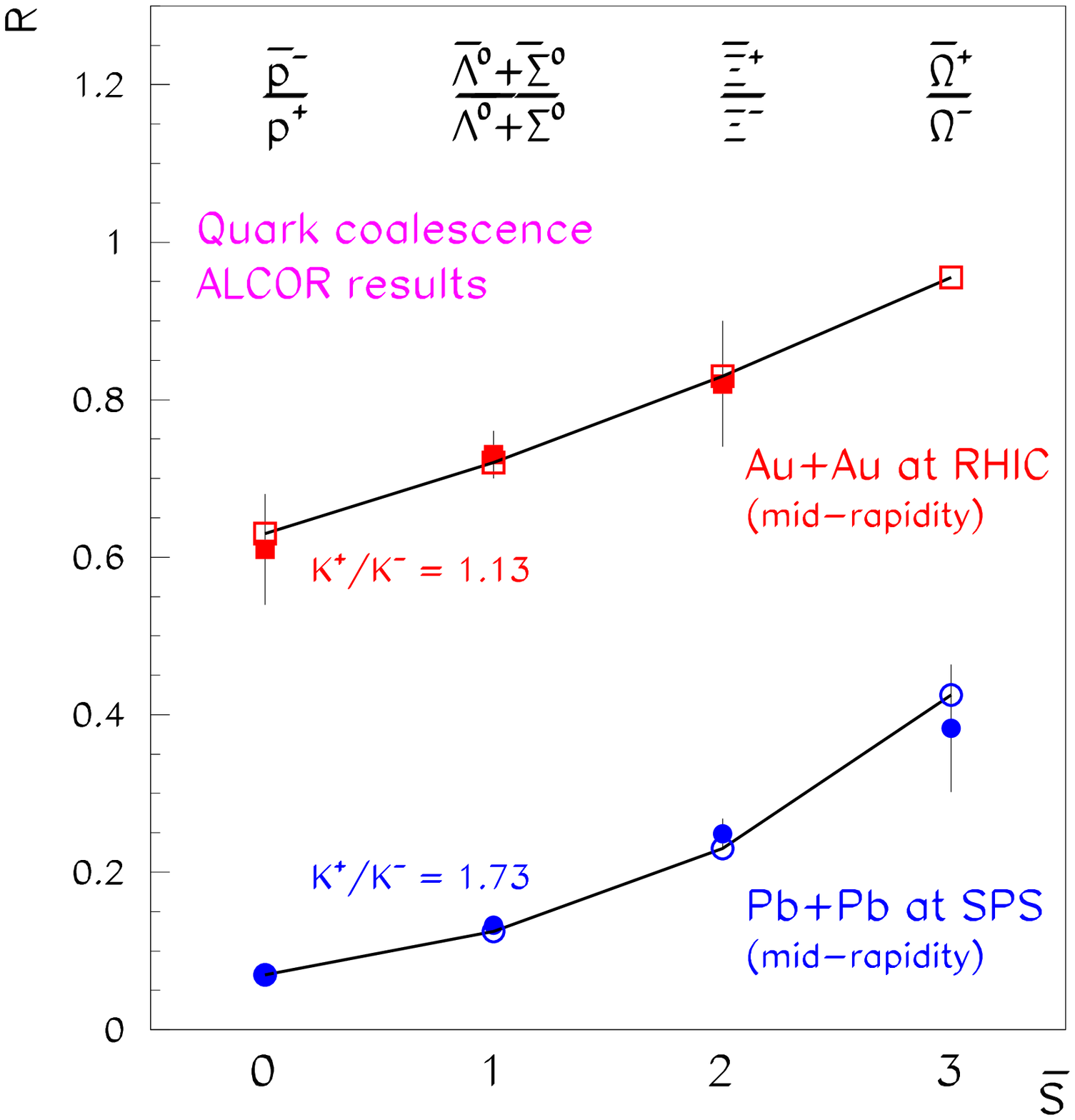}
\vspace{-4.2cm}
\end{center}
\begin{center}
\begin{minipage}[t]{11.5cm}
         {\bf Fig. 4.} {\small 
Calculated (open symbols) and measured (full symbols) hyperon
 ratios for SPS (WA97) and RHIC (STAR) experiments, see Ref.~\cite{QM99}.}
\end{minipage}
\end{center}

\section{ Energy dependence of physical quantities}

Since the rapidity density of charged particles is increasing
with increasing energy, we can investigate the energy dependence
of the ALCOR parameters. From available AGS, SPS and RHIC data
(see Fig. 5), one can determine the 3 necessary parameters and
calculate the results from the ALCOR model. Fig. 6. displays
the tendency of the stopping into the mid-rapidity region
and Fig. 7 shows the energy dependence of the
number of newly produced light quark-antiquark pairs ($dN_{u{\overline u}}/dy$)
and the strange ones indicated by $f_s$ in a logarithmic energy scale.

\begin{center}
\vspace*{10.7cm}
\includegraphics{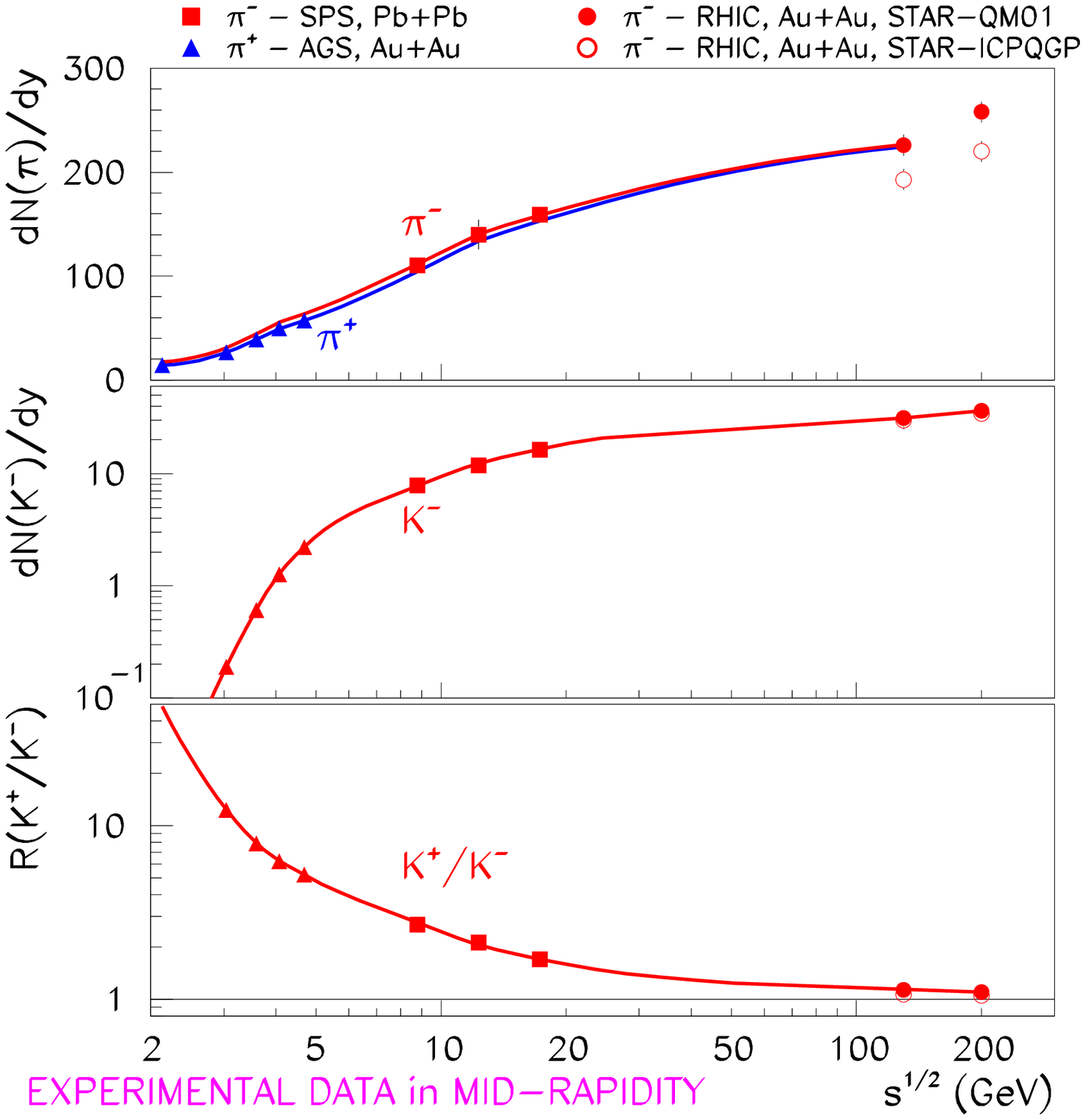}
\vspace{-4.2cm}
\end{center}
\begin{center}
\begin{minipage}[t]{11.5cm}
         {\bf Fig. 5.} {\small 
The three input (measured) data  for ALCOR as a function of energy.
Data are from the AGS, SPS and RHIC experiments.}
\end{minipage}
\end{center}

\begin{center}
\vspace*{9.7cm}
\includegraphics{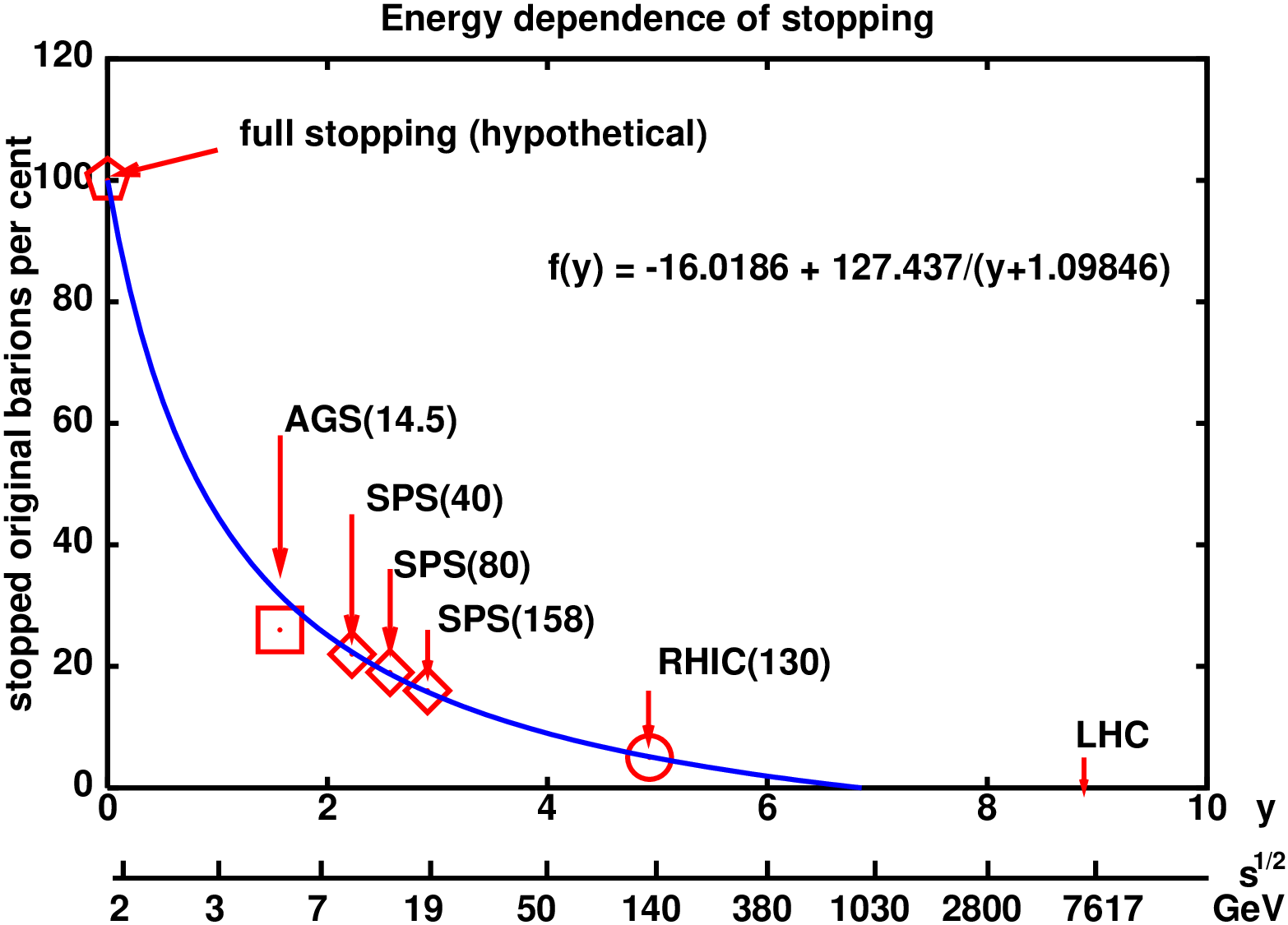}
\vspace{-2.2cm}
\end{center}
\begin{center}
\begin{minipage}[t]{11.5cm}
         {\bf Fig. 6.} {\small 
Energy dependence of the stopping parameter in the mid-rapidity region. }
\end{minipage}
\end{center}

\newpage

\begin{center}
\vspace*{8.7cm}
\includegraphics{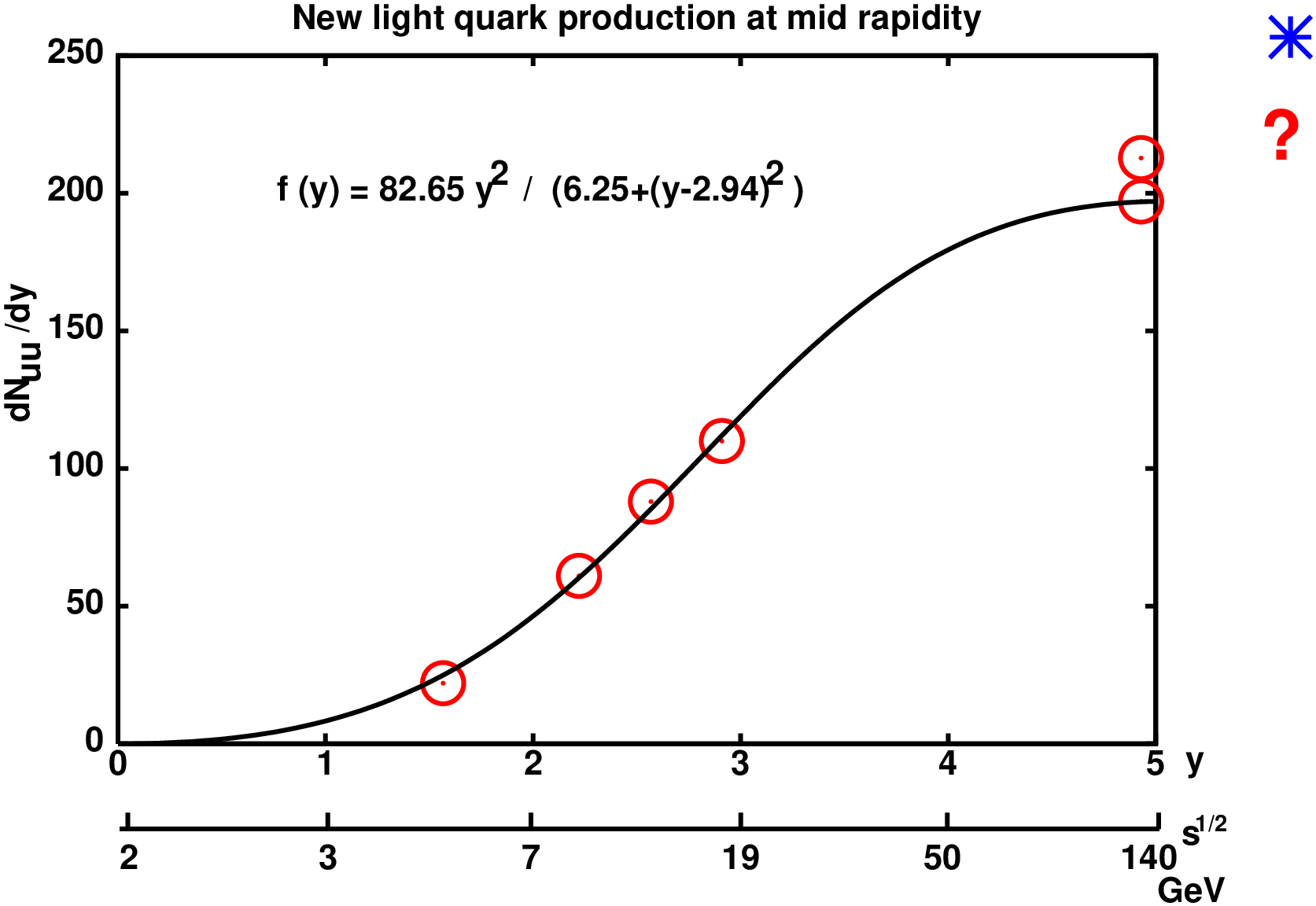}
 
\includegraphics{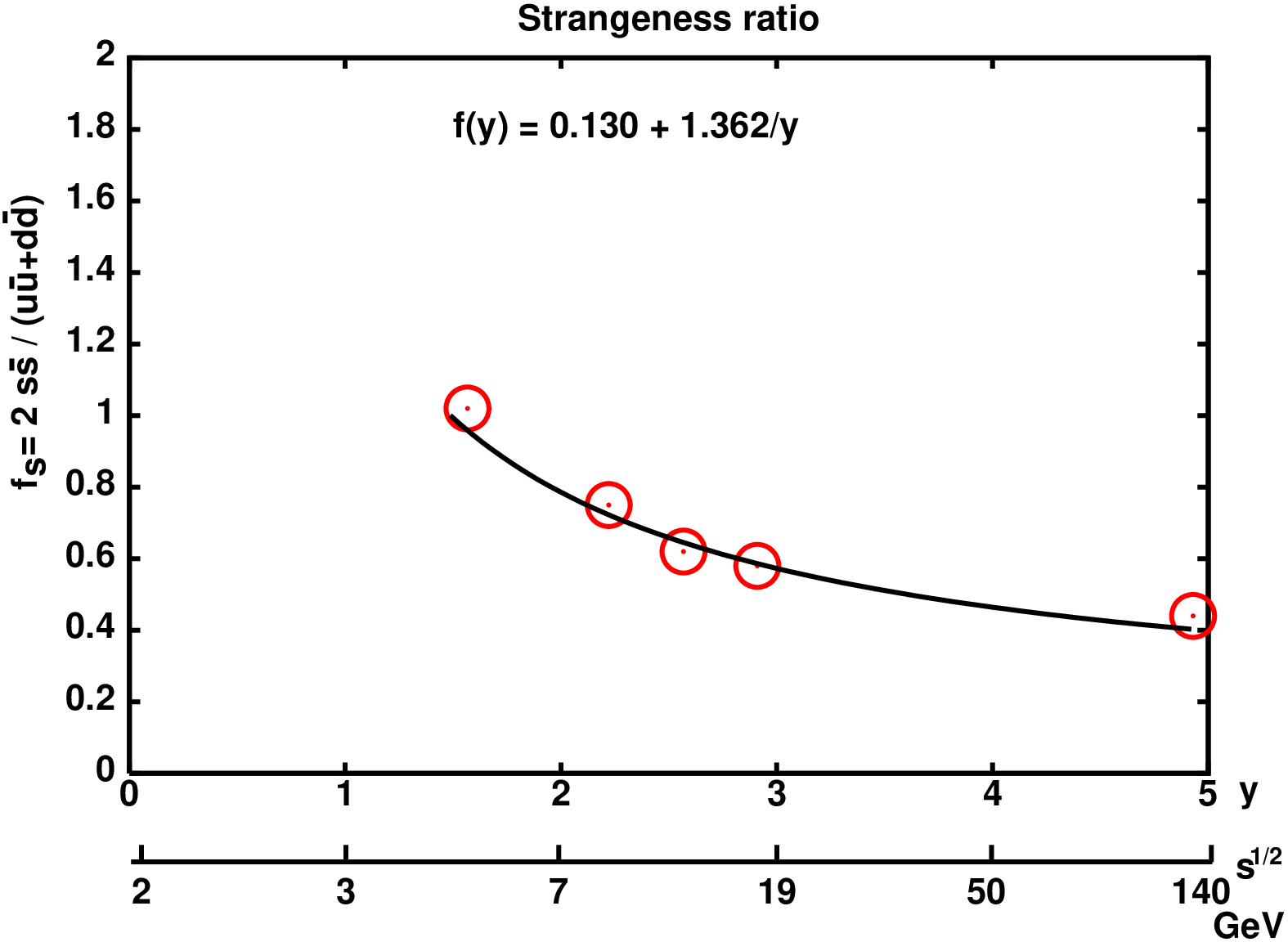}
 
\vspace{-3.5cm}
\end{center}
\begin{center}
\begin{minipage}[t]{11.5cm}
         {\bf Fig. 7.} {\small  
The energy dependence of the ALCOR parameters:
the rapidity density, $dN_{uu}/dy=dN_{dd}/dy$, of newly
produced light quark-antiquark pairs (left panel);
ratio of the newly produced strange quark-antiquark quarks
to non-strange ones, $f_s$ (right panel).}
\end{minipage}
\end{center}

\section{Charge fluctuation}

In a recent paper \cite{Bial3} interesting observation was
 made on the problem of 
charge fluctuation. In earlier papers it was suggested
\cite{Heinz}, that the ratio
\begin{equation}
    D = 4 \frac{< \delta Q^2 >}{< N_{ch} >}
\end{equation}
is  $ D \approx 3 $ for hadron gas in equilibrium, while
it is  $ D \approx 1 $ for the quark gluon plasma.
Here $ < \delta Q^2 > $ is the average of the charge fluctuation
and $ < N_{ch} > $ is the average value of the charge multiplicity.
The measured $ D $ value is near to 3 . That would indicate 
that in heavy ion reactions not a quark-gluon plasma,
but a hadron gas was produced. However, in
  Ref.\cite{Bial3},
it was shown, that in the coalescence scenario (ALCOR) one can
also expect a $ D \approx 3 $ value.
Thus one can conclude that the measured charged fluctuation 
is also in agreement with the basic assumption of the ALCOR,
i.e. during the hadronization the number of constituent quarks and 
the number of constituent antiquarks are conserved.

\section{ Conclusion}

In this paper we have shown that the assumptions of the ALCOR model
are in agreement with the experimental facts. These
assumptions are:
a) the matter just before the hadronization consists of a mixture
of massive quarks and antiquarks; b) the number both of the quarks 
and the antiquarks are conserved during the sudden hadronization; 
c) the hadrons are formed by  coalescence process. 

We have investigated the energy dependence of particle ratios and
thus found that the Wroblinski factor ($f_s$)  has a definite
energy dependence. This indicates that the strange quark-antiquark 
pairs and the light quark-antiquark pairs are produced 
with different mechanisms having different energy dependence.

We quoted Ref.~\cite{Bial3}, where it was shown that 
the charge fluctuations calculated from the massive quark matter
is also in agreement with the experimental results.

Finally we may conclude, that  we have not seen the
signature of  an ideal quark-gluon plasma, but {\bf we found that in the heavy 
ion collisions a new type of matter, the massive 
quark-antiquark matter is produced}. 
\bigskip

{\bf Note added in proof:}
The quark coalescence model was misunderstood many times and it induced
surprisingly strong misinterpretations. This situation is well demonstrated 
in the following criticism~\cite{Hwa}:
 
 "The emphases in refs. \cite{Bial98,Zim99} are in the multiplicative aspect of
the probabilities of having quarks and antiquarks in the same region of
the phase space in their formation of hadrons. That results in an
undesirable feature of $s$ and {$\overline s$ } imbalance in the linear
version \cite{Bial98}, which is not satisfactorily resolved in the
nonlinear version {\cite{Zim99} } by the introduction of unknown
factors."

  On the contrary, in Refs.~\cite{Bial98,Zim99} particle ratios are
calculated in a way where all unknown factors drop from the results. In
the nonlinear version the $ b_i $ coefficients are normalization factors
ensuring the different particle conservation rules. To interpret these
normalization factors as "the introduction of unknown factors" is equivalent
to stating that conservation of e.g. electric charge is an unknown rule.

\section*{Acknowledgments}

This work was supported by the grant of 
Hungarian Scientific Research Found OTKA T034269.

\end{document}